# X-ray photoemission study of CoFeB/MgO thin film bi-layers

J.C. Read, P.G. Mather[a], and R.A. Buhrman

School of Applied and Engineering Physics

Cornell University, Ithaca, NY 14853-2501

**ABSTRACT**

We present results from an X-ray photoemission spectroscopy (XPS) study of CoFeB/MgO bi-layers where we observe process-dependent formation of B, Fe, and Co oxides at the CoFeB/MgO interface due to oxidation of CoFeB during MgO deposition. Vacuum annealing reduces the Co and Fe oxides but further incorporates B into the MgO forming a composite $MgB_xO_y$ layer. Inserting an Mg layer between CoFeB and MgO introduces an oxygen sink, providing increased control over B content in the barrier.

---

[a] Current address: Freescale Semiconductor, Chandler, AZ 85224

Magnetic tunnel junction (MTJ) devices formed by insertion of a thin insulating layer between ferromagnetic electrodes provide a means to form memory elements for magnetic random access memory and read head sensors for high-density data storage applications. To achieve such magnetoelectronic technology, MTJ structures with large tunneling magnetoresistance (TMR) and an appropriately impedance-matched device resistance-area (RA) product are required. Recently it was demonstrated that MgO-based MTJs utilizing the electrode material CoFeB and radio frequency (rf) sputtering to deposit the MgO tunnel barriers can exhibit very large TMR values, [1,2] indicating that careful engineering of the tunnel barrier can yield significant improvements in device performance and versatility. In addition the inclusion of a thin Mg layer between the bottom electrode and a thin MgO barrier layer can yield devices with a low RA product and a high TMR.[3,4] These advances motivate investigation of the chemistry and electronic properties of the MgO barrier layer and the electrode/MgO interface with the goal of developing pathways for still further improvements, particularly for ultra-thin barriers.

We present results from an XPS study of CoFeB/MgO bi-layers in which we observe the formation of a composite Mg and B oxide layer. While B oxide formation is more pronounced in rf sputtered MgO layers evaporated MgO layers also show some oxidic B. Annealing in vacuum at 375 C increases the B concentration in the MgO layers as B diffuses from the electrode into the oxide. We also observe the formation of interfacial Fe and Co oxides for thicker rf sputtered MgO layers but these oxides are substantially reduced after annealing. Deposition of a thin Mg layer on the electrode before rf deposition of MgO inhibits the formation of interfacial oxides and forms an MgO barrier similar to evaporated MgO.

The thin film stacks were grown on thermally oxidized Si(100) substrates in a vacuum system with a base pressure of ~ 2 x $10^{-9}$ Torr. The 20 nm base electrodes were deposited using dc magnetron sputtering in Ar (1 x $10^{-3}$ Torr) from an alloy target ($Co_{60}Fe_{20}B_{20}$). We used several deposition techniques to form the oxide layers: (1) electron-beam evaporation of stoichiometric source material (MgO[eb]); (2) rf sputtering of a stoichiometric target in Ar (1 x $10^{-3}$ Torr) (MgO[rf]); and (3) sputtering of bi-layer structures consisting of a thin (0.5 nm or 1.0 nm) dc sputtered Mg layer followed by an MgO[rf] layer. Sample structures were prepared in identical pairs with one of each pair receiving vacuum annealing at 375 C for one hour.

Once prepared we vacuum-transferred samples to a Surface Science Laboratories SSX-100 x-ray photoelectron spectrometer employing monochromatic Al $K_\alpha$ x-rays (1487 eV) with the pass energy of the hemispherical detector kept at 25 eV and a system resolution of approximately 0.5 eV. We calibrated the detector to the Au $4f_{7/2}$ line at 84 eV and sample spectra to the strong intensity metallic $Co2p_{3/2}$ and $Fe2p_{3/2}$ lines (not shown) at 778.3 and 707.0 eV respectively.[5]

Fig. 1 shows B 1s, Mg 2p, and O 1s XPS spectra from MgO[eb] samples. The B 1s spectra (Fig. 1a) exhibit a peak at ~ 188 eV and a small peak ~ 192 eV due to B in the electrode and oxidized B respectively.[6,7] Fig. 1b shows the Mg 2p, Fe 3p, and Co 3p spectral region. The metallic Co 3p (~ 59.4 eV) and Fe 3p (~ 52.7 eV) peaks[5] provide comparison for the peak at ~ 50 eV due to Mg cations in the MgO layer.[8,9] These peaks also allow investigation of the presence of Co (~ 61 eV) and Fe (~ 55-56 eV) oxides at the CoFeB/MgO interface, which we confirm by study of the Co 2p and Fe 2p spectra. There is no indication of Co or Fe oxide in these MgO[eb] samples, and we attribute the slight oxidation of B at the electrode surface to O ions liberated during evaporation.

Fig. 2 shows XPS spectra from MgO[rf] samples. The $BO_x$ peak at ~ 192 eV is quite large compared to MgO[eb] samples and the relative intensity of the $BO_x$ peak is greater for a 2 nm MgO[rf] sample where $FeO_x$ and $CoO_x$ are also clearly present. This establishes that $BO_x$ is formed throughout the rf process by oxygen ions released by the sputtering of the MgO target. The observation that the initial formation of $BO_x$ is favored in electrode oxidation is consistent with XPS measurements of CoFeB electrodes exposed to different $O_2$ doses. There a boron rich oxide forms initially, while Fe and then Co oxides develop later, upon exposure of the electrode to greater doses.

Fig. 3 shows XPS spectra from CoFeB/Mg/MgO[rf] tri-layer samples. There is no clear indication of $BO_x$ prior to annealing when a 1 nm thick Mg layer is inserted between the CoFeB and a 0.5 nm MgO[rf] layer. In comparison the 0.5 nm Mg/ 1 nm MgO[rf] bi-layer displays a substantial $BO_x$ signal. We conclude that the underlying Mg layer serves effectively as an O sink (getter) during the rf sputter process until fully consumed, after which preferential formation of $BO_x$ at the CoFeB/MgO interface begins.

The O 1s spectra for MgO[eb] samples in Fig. 1c display two distinct peaks, one at ~ 530 eV attributable to O in the MgO layer, the other at ~ 533 eV, attributable to chemisorbed oxygen and possibly hydroxide species on the MgO surface.[8,9] Both the O 1s and the Mg 2p peaks shift slightly to higher binding energy (BE) as the film thickness increases from 1 to 2 nm, which we attribute to the image charge effect[10]. Compositional analysis using the lower BE O 1s and Mg peaks finds these MgO[eb] layers are approximately stoichiometric.

The O 1s spectra from Mg/MgO[rf] samples in Fig. 3c provide insights for understanding the MgO[rf] O 1s spectra. The O 1s spectrum from the 1 nm Mg/ 0.5 nm MgO sample is similar to spectra from MgO[eb] samples, and compositional analysis using the low

BE O 1s peak indicates that the 1 nm Mg/ 0.5 nm MgO[rf] process results in the formation of an MgO layer that is approximately stoichiometric. In comparison, while the 0.5 nm Mg/1nm MgO[rf] sample also shows an O 1s peak at ~ 530 eV, the higher BE O 1s peak is larger and centered at ~ 532 eV while the B 1s spectrum indicates a substantial oxidic B component. Thus we attribute the large and broad O 1s peak at ~ 532 eV for this sample to a mixture of surface oxygen and $BO_x$.[6,7,11]

The O 1s spectrum from the 1 nm MgO[rf] sample shown in Fig. 2c also has a broad dominant peak at ~ 531 eV and a much smaller peak at ~ 530 eV, with the latter attributable to $FeO_x$. Based on the previous peak identifications and the clear presence of significant $BO_x$ we attribute the higher BE peak to $BO_x$, surface oxygen, and possibly some mixed oxide $MgB_xO_y$, as discussed below. The O 1s spectrum for the 2 nm MgO[rf] sample can also be fit with two broad peaks, but both peaks are shifted to higher BE. The lower BE peak has gained intensity and is likely the convolution of MgO, $FeO_x$, and $CoO_x$ signals.

Compositional analysis of all the as-grown CoFeB/MgO samples shows some B incorporation in the oxide, with the MgO[eb] layers containing < 2% B and the MgO[rf] layers substantially more. The B:Mg cation ratio in the oxide is ~ 0.63 for the 1 nm MgO[rf] sample and 0.27 for the 2 nm MgO[rf] sample. The decreased B:Mg ratio with increased film thickness indicates that the $BO_x$ is formed more readily during the initial stages of the MgO deposition process but the $BO_x$ signal increase with increased MgO thickness indicates that some electrode oxidation takes place throughout the deposition process. For the Mg/MgO bilayers there is no clear signal of $BO_x$ in the 1 nm Mg/ 0.5 nm MgO[rf] sample while the B:Mg ratio in the 0.5 nm Mg/ 1 nm MgO[rf] oxide is ~ 0.25, similar to the ratio for the 2 nm MgO[rf] layer.

Annealing is necessary to achieve the highest TMR in CoFeB/MgO/CoFeB MTJs.[1-4] Figs. 1-3 show the affect of annealing upon our CoFeB/MgO samples. For MgO[eb] samples (Fig. 1) the $BO_x$ peak at ~ 192 eV increases slightly indicating that B cations diffuse into the oxide during annealing. However compositional analysis, which for simplicity assumes uniform distribution, finds that the maximum B concentration in the MgO[eb] oxide remains < 3 %. The effects of annealing are considerably different for MgO[rf] samples (Fig.2). The rf sputtering process forms substantial $BO_x$ and annealing further increases the intensity of the $BO_x$ signal in MgO[rf] films with the B:Mg cation ratio in the oxide changing from 0.6 to 0.8 after annealing for a 1 nm MgO[rf] layer and from 0.3 to 0.5 for a 2 nm MgO[rf] layer. Again this indicates that a very substantial amount of additional B is diffusing into the MgO barrier during annealing. XPS depth profiling studies have yielded a similar conclusion for annealed CoFeB/MgO-based structures.[12] We also note that annealing significantly reduces the Co and Fe oxide peaks and shifts the $FeO_x$ peak to lower BE, indicating a reduction in oxidation state. Apparently B from the electrode is consuming O liberated by reduction of the Co and Fe oxides.

An important point to note is that annealing shifts the oxidic Mg and B peaks to higher BE by roughly 1 eV for the 1 nm thick MgO[rf] layer. A shift of almost 1 eV is also apparent in the O 1s spectrum and the broad peak structure has become more uniform after annealing. Peak shifting is also present in 2 nm MgO[rf] and 0.5 nm Mg/1.0 nm MgO samples with, in general, the degree of oxide peak shift scaling with the pre-anneal oxidic B concentration. This collective shifting of the Mg, B, and O peaks in samples with significant $BO_x$ content suggests that after annealing the oxide is an intermixed $MgB_xO_y$ layer. In this case the upward shifts of the Mg, B, and O peaks can be explained by the $Mg^{+2}$ cations being in a lower average state of O coordination than in pure MgO and the oxidic B in a more ordered or a higher oxidation ($B^{+3}$ as opposed to $B^{+2}$) state after annealing.[13]

In contrast, the spectra for the 1 nm Mg/ 0.5 nm MgO[rf] sample (Fig. 3) show that, when there is little $BO_x$ initially present, annealing, while it measurably increases the oxidic B signal, does not have the effect of making B a substantial component (> 5 %) of the oxide. After annealing the B:Mg cation ratio in the oxide becomes ~ 0.1 for the 1 nm Mg/ 0.5 nm MgO[rf] sample and changes from 0.25 to 0.6 for the 0.5 nm Mg/ 1 nm MgO[rf] sample. Thus if the initial $BO_x$ content is low, after annealing there is no detectable shift of the Mg 2p peak and the oxide has the character of MgO that is doped with B. If the oxidic B content is higher, a major shift of the Mg, B, and O peaks to higher BE occurs upon annealing suggesting the formation of an atomically mixed $MgB_xO_y$ barrier.

Finally we note that the lack of $FeO_x$ and $CoO_x$ in the MgO[eb] samples demonstrates that significant oxidation of CoFeB does not take place during our dc sputtering processes or upon limited exposure to typical vacuum system pressures. Instead it is the deposition of MgO that causes oxidation of the electrode. Thus we attribute the increased control and quality of MTJs with MgO[rf] barriers that is achieved by inclusion of an Mg layer,[3,4] and by deposition of a Ta gettering layer before MgO[rf] deposition,[14,15] to the capture of oxygen liberated from the MgO target in the early stages of sputtering.

In summary, using XPS we observe the formation of $BO_x$ and in some cases $FeO_x$ and $CoO_x$ at the interface of sputter-deposited CoFeB/MgO[rf] bi-layers. We attribute this result to the evolution of substantial oxygen ions in the deposition of MgO. Vacuum annealing at 375 C promotes reduction of the $FeO_x$ and $CoO_x$ and forms a mixed $MgB_xO_y$ layer. Both e-beam deposition of MgO and insertion of a thin Mg layer between the CoFeB electrode and the MgO layer substantially reduces or eliminates the amount of B, Fe, and Co oxides initially formed. but we invariably find that some B is still incorporated into the MgO tunnel barrier after annealing. While the correlation of MTJ performance with barrier composition is beyond the scope of this initial XPS investigation, this work demonstrates the complexity of the barriers that can form, depending upon the details of the MgO process. We note that some incorporation of B into the MgO may be quite beneficial particularly if it acts to fill defect sites in the oxide. A recent report discusses MTJs with ultra-low RA and high TMR[14] obtained with the use of Ta as a getter just prior to MgO[rf] deposition. An XPS study[15] of the barrier material yielded an O 1s spectrum fully consistent with a significant $BO_x$ component. Without the Ta getter the high BE O 1s peak intensity is much larger, suggesting that while a large B concentration that changes the crystal structure of a nominally MgO tunnel barrier may not be beneficial for TMR, some B incorporation may well be helpful.

This research was supported by the NSF MRSEC program through the Cornell Center for Materials Research, by ONR and by NSF through use of the facilities of the Center for Nanoscale Systems.

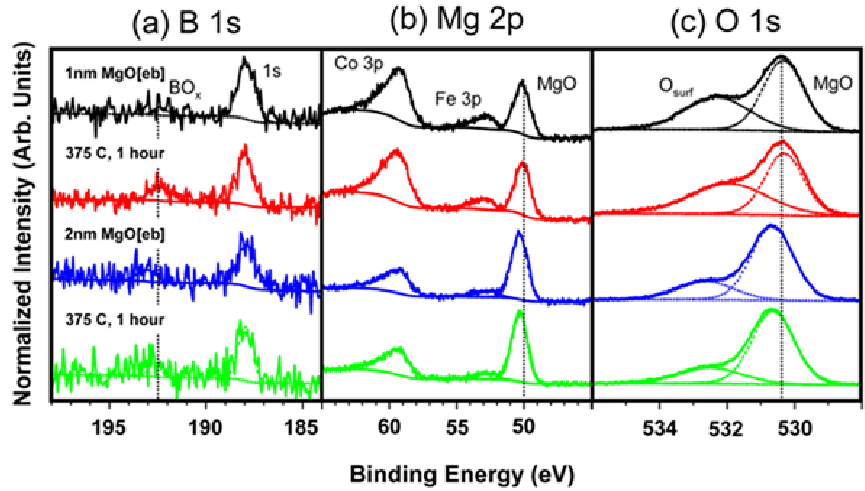

Figure 1, Read *et al.*

Figure 1. Normalized XPS spectra from the B 1s (a), Mg 2p (b), and O 1s (c) spectral regions for CoFeB/MgO[eb] samples before and after annealing at 375 C. Spectra are normalized to their individual maximum peak intensities in each spectral region and offset for clarity. The $BO_x$ signal increases after annealing as B diffuses into the MgO.

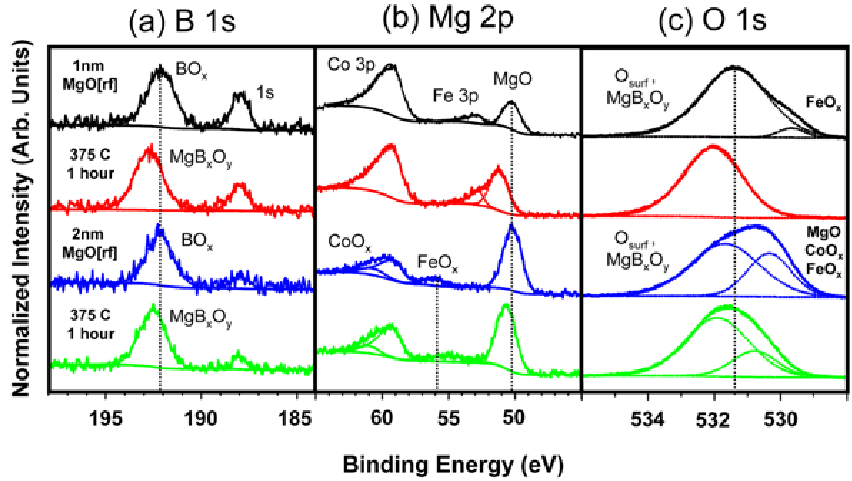

Figure 2, Read *et al.*

Figure 2. Normalized XPS spectra for CoFeB/MgO[rf] samples before and after annealing. Significant $BO_x$ is present in all samples, and the $BO_x$, MgO (Mg 2p) and O 1s peaks shift to higher BE after annealing, indicating formation of $MgB_xO_y$. $FeO_x$ and $CoO_x$ form in 2 nm thick MgO barriers but both oxides are reduced by annealing.

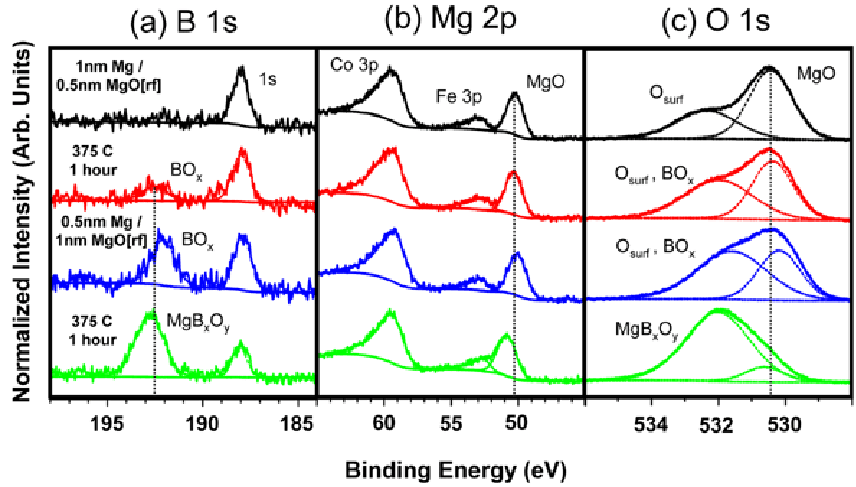

Figure 3, Read *et al.*

Figure 3. Normalized XPS spectra for CoFeB/Mg/MgO[rf] samples before and after annealing. An Mg layer between CoFeB and MgO[rf] suppresses $BO_x$, $FeO_x$, and $CoO_x$ formation but annealing promotes B diffusion into the MgO tunnel barrier. The O 1s peak at ~ 532 eV grows in intensity with the $BO_x$ peak providing evidence both are characteristic of B content in the MgO tunnel barrier.